\documentstyle[amssymb,aps,prb,12pt]{revtex}

\begin{document}
\title{Propagation of axions in a strongly magnetized medium\thanks{%
Revised version of the paper published in J. Exp. Theor. Phys. {\bf 88}, 1
(1999).}}
\author{A. V. Borisov\thanks{%
E-mail: borisov@ave.phys.msu.su} and P. E. Sizin}
\address{M. V. Lomonosov Moscow State University, 119899 Moscow, \\
Russia}

\maketitle

\begin{abstract}
The polarization operator of an axion in a degenerate gas of electrons
occupying the ground-state Landau level in a superstrong magnetic field $%
H\gg H_0=m_e^2c^3/e\hbar =4.41\cdot 10^{13}$ G is investigated in a model
with a tree-level axion-electron coupling. It is shown that a dynamic axion
mass, which can fall within the allowed range of values $(10^{-5}$ eV $%
\lesssim m_a\lesssim 10^{-2}$ eV$)$, is generated under the conditions of
strongly magnetized neutron stars. As a result, the dispersion relation for
axions is appreciably different from that in a vacuum.

\vspace{0.25in}
\end{abstract}


{\bf 1}. The {\it a priori} strong nonconservation of $CP$ parity in the
standard model can be eliminated in a natural manner by introducing axions
--- pseudo-Goldstone bosons associated with the spontaneous breaking of the
additional Peccei--Quinn global symmetry $U(1)_{PQ}$.\cite{r1,r2} According
to the experimental data,\cite{r3} the energy scale $v_a$ for the $U(1)_{PQ}$
symmetry breaking is much greater than the electroweak scale --- $v_a\gtrsim
10^{10}$ GeV, and the constants of the possible couplings of an axion to the
standard particles $(\sim 1/v_a)$ are very small (the ``invisible'' axion:
see Ref. \onlinecite{r4} for a review of various axion models).

Axion effects can be appreciable under the astrophysical conditions of high
matter densities, high temperatures, and strong magnetic fields (for
example, in neutron stars\cite{r5}). Axion production processes, which
result in additional energy losses by stars, and the limits obtained by
astrophysical methods on the parameters of axion models are examined in Ref. %
\onlinecite{r4}. In so doing, the influence of electromagnetic fields was
neglected.

The investigation of axion processes in strong magnetic fields commenced
comparatively recently. The Compton and Primakoff mechanisms of axion
production on nonrelativistic electrons by thermal photons $(\gamma
+e\rightarrow e+a)\,$ in the presence of a magnetic field are studied in
Ref. \onlinecite{r6}. The extension to relativistic electrons in a constant
external electromagnetic field is given in Ref. \onlinecite{r7} (Primakoff
effect) and Refs. \onlinecite{r8} and \onlinecite{r9} (Compton effect), where%
\cite{r7,r9} estimates were also obtained for the contributions of the
indicated processes to the axion luminosity of a magnetized strongly
degenerate relativistic electron gas under the conditions of the crust of a
neutron star. A new axion production mechanism --- synchrotron emission of
axions $(e\rightarrow e+a)\,$ by relativistic electrons --- was proposed in
Ref. \onlinecite{r10} and its contribution to the energy losses by a neutron
star was calculated. In Refs. \onlinecite{r6} --\onlinecite{r10} it was
assumed that the external field intensity $F\ll H_0=m_0^2c^3/e\hbar \simeq
4.41\cdot 10^{13}$ G. In Ref. \onlinecite{r11}, numerical methods were used
to extend the results of Ref. \onlinecite{r10} to superstrong magnetic
fields $H\gtrsim H_0$. It was found that the basic equation derived in Ref. %
\onlinecite{r10} for the axion synchrotron luminosity for the semiclassical
case of high electron energies $(\varepsilon \gg m_ec^2)\,$ and fields $H\ll
H_0$ agrees with the numerical calculations up to $H/H_0\lesssim 20$. The
axion synchrotron luminosity of neutron stars and white dwarfs was also
investigated in Ref. \onlinecite{r11}.

In Refs. \onlinecite{r8} --\onlinecite{r11} a model with a derivative
axion-electron coupling $eae$, described by the interaction Lagrangian\cite
{r4}

\begin{equation}
{\cal L}_{ae}=\frac{g_{ae}}{2m_e}\left( \overline{\psi }\gamma ^\mu \gamma
^5\psi \right) \partial _\mu a,
\end{equation}
was used. Here $m_e$ is the electron mass and $\gamma ^5=-i\gamma ^0\gamma
^1\gamma ^2\gamma ^3$; the system of units such that $\hbar =c=1$ is used;
the signature of the metric is $(+---)$; and

\begin{equation}
g_{ae}=c_e\frac{m_e}{v_a}
\end{equation}
is a dimensionless coupling constant, where the numerical factor $c_e$
depends on the choice of the specific axion model.\cite{r4}

In models where axions are coupled only with heavy fermions by a tree-level
coupling there arises an effective direct low-energy axion--photon
interaction of the type $\gamma a\gamma $.\cite{r4} This interaction is the
basis of the Primakoff axion photoproduction mechanism employed in Refs. %
\onlinecite{r6} and \onlinecite{r7}. The synchrotron process $e\rightarrow
ea\,$ in the absence of a tree-level axion--electron coupling was considered
recently in Ref. \onlinecite{r12}. This process is due to resonant
conversion of a longitudinal plasmon (a photon in a medium), emitted by a
relativistic electron in a magnetic field, into an axion.

Decay of an axion in a strong magnetic field into a fermion pair $%
(a\rightarrow f\bar{f})$\cite{r13} and two photons $(a\rightarrow \gamma
\gamma )$\cite{r14} are also of interest for astrophysics and cosmology.

In the present paper the model (1) is used to calculate the polarization
operator of an axion moving in a strongly magnetized degenerate electron gas
and the change in the dispersion relation of an axion in a medium is
investigated using this operator.

{\bf 2}. Taking account of the contribution of the electrons only (see Eq.
(1)) we obtain, using the real-time formalism of the finite-temperature
quantum field theory (see, for example, Ref. \onlinecite{r15}), the
following momentum representation for the one-loop polarization operator of
an axion:

\begin{equation}
\Pi (k,k^{\prime })=-iG_a^2\int d^4x\,d^4x^{\prime }\exp (ikx- ik^{\prime
}x^{\prime }){\rm Tr}\left[ \hat{k}\gamma ^5G(x,x^{\prime })\hat{k}^{\prime
}\gamma ^5G(x^{\prime },x)\right] .
\end{equation}
Here $k$ $(k^{\prime })$ is the final (initial) 4-momentum of an axion; $%
G(x,x^{\prime })$ is the time-dependent single-particle Green's function of
an ideal electron-positron gas in a constant magnetic field;\cite{r15}
notations have also been introduced for the contraction $\hat{a}=\gamma ^\mu
a_\mu $ of a 4-vector $a^{\mu \text{ }}$ with the Dirac $\gamma $ matrices
and for the dimensional coupling constant

\begin{equation}
G_a=\frac{g_{ae}}{2m_e}.
\end{equation}
On account of the translational invariance (constant external field,
homogeneous isotropic medium) the polarization operator (3) is diagonal in
momentum space:

\begin{equation}
\Pi (k,k^{\prime })=(2\pi )^4\delta ^{(4)}(k-k^{\prime })\Pi (k).
\end{equation}
Here $\Pi (k)$ determines the axion propagator $D(k)$ in the momentum
representation according to the Dyson equation

\begin{equation}
D(k)=\left[ k^2-m_a^2-\Pi (k)\right] ^{-1},
\end{equation}
where $m_a$ is the free-axion mass (in the absence of a field and a medium),
which is generated by the chiral anomaly of QCD:\cite{r2} $m_a\sim \Lambda
_{QCD}^2/v_a$. The renormalized value $\Pi _R(k)$ (see below) gives the
dispersion relation

\begin{equation}
k^2=m_a^2+\Pi _R(k).
\end{equation}

{\bf 3}. We give the constant uniform magnetic field ${\bf H\,\Vert \,\hat z}
$ in terms of the 4-potential $A^\mu $ in the gauge

\begin{equation}
A^\mu =(0,0,xH,0).
\end{equation}
Then the Green's function $G(x,x^{\prime })$ can be represented in the
following form after summing over the spin quantum number and the sign of
the energy in the general expression for $G\,$ in the form of a series in
quadratic combinations of the eigenfunctions of the Dirac operator\cite{r15}:

\[
G(x,x^{\prime})=\left[\gamma^\mu(i\partial _\mu+eA_\mu)+m_e
\right]K(x,x^{\prime});
\]
\[
K(x,x^{\prime})=\frac{\sqrt{h}\,}{(2\pi)^3}\, \sum \limits_{n=0}^\infty
\int\limits_{-\infty}^{\infty} dp_0dp_ydp_z\exp \left[
-ip_0(t-t^{\prime})+ip_y(y-y^{\prime})+ip_z(z-z^{\prime}) \right]\times
\]
\begin{equation}
\times u_n( \eta)u_n(\eta^{\prime})\left( R_{n+1}\Sigma_++R_n\Sigma_-
\right),
\end{equation}
\[
R_n= \left[p_0^2-p^2_z-2hn-m^2_e+i0 \right]^{-1}+ 2\pi i
\delta(p_0^2-p^2_z-2hn-m^2_e)N_F(p_0).
\]

Here the electron charge $-e<0,$ $h=eH;$ $n=0,\,1,$ $2,\,...$ is the
principal quantum number (the number of the Landau level); $p_y$ and $p_z$
are the eigenvalues of the operators of the canonical momenta --- the constants of motion in the gauge (8); and $u_n(\eta )$ is a Hermite
function of argument

\[
\eta =\sqrt{h}(x+p_{y}/h);\quad \eta ^{\prime }=\eta (x\rightarrow x^{\prime
});\quad \Sigma _{\pm }=\left( 1\pm \Sigma _{3}\right) /2,\quad \Sigma
_{3}=i\gamma ^{1}\gamma ^{2}.
\]

The first term in $R_n$ has poles at the points $p_0=\pm \varepsilon =\pm
[m_e^2+2hn+p_z^2]^{1/2}$, determining the energy spectrum of an electron in
a magnetic field. The second term $(\propto \delta (p_0^2-\varepsilon ^2))$
describes the effect of the electron-positron medium, and

\begin{equation}
N_F(p_0)=\theta (p_0)\left[ \exp \left[ \beta (p_0-\mu )\right] +1\right]
^{-1}+\theta (-p_0)\left[ \exp \left[ \beta (-p_0+\mu )\right] +1\right]
^{-1}
\end{equation}
is expressed in terms of the Fermi distribution function of electrons and
positrons in a medium with temperature $T=1/\beta $ and chemical potential $%
\mu $, and $\theta (\pm p_0)\,$ is the Heaviside step function.

{\bf 4}. It is difficult to make a general analysis of the axion
polarization operator for arbitrary values of the parameters $H,$ $T,$ and $%
\mu $. In the present paper we confine our attention to superstrong magnetic
fields and comparatively low temperatures

\begin{equation}
H\gg H_0,\quad T\ll \mu -m_e,
\end{equation}
and we require the chemical potential to satisfy
\begin{equation}
\mu ^2-m_e^2<2h.
\end{equation}
It follows from Eqs. (11) and (12) that in this case the contribution of
positrons in Eq. (10) can be neglected (it is suppressed by the factor $\exp
[-\beta (\mu -m_e)]$) and the medium is a degenerate gas of electrons
occupying the ground-state Landau level $(n=0)$:

\begin{equation}
N_F(p_0)=\theta (p_0)\theta (\mu -p_0),\quad p_0=\sqrt{m_e^2+p_z^2}.
\end{equation}
We also limit the range of the axion 4-momentum

\begin{equation}
|k_0^2-k_z^2|\ll h.
\end{equation}
Then the main contribution of virtual (vacuum) electrons and positrons is
likewise formed by states with $n=0$. As a result, retaining on the basis of
Eqs. (11), (12), and (14) terms with $n=0$ in the sum (9), we obtain the
following approximate expression for the Green's function in a superstrong
magnetic field:

\begin{eqnarray}
G(x,x^{\prime }) &\simeq &\left( \frac h\pi \right)
^{1/2}\int\limits_{-\infty }^\infty \frac{dp_y}{2\pi }\exp \left[ - \frac 12%
(\eta ^2+\eta ^{\prime }{}^2)+ip_y(y-y^{\prime })\right] \times  \nonumber \\
&\times &\int \frac{d^2p}{(2\pi )^2}\exp \left[ -ip_0(t-t^{\prime
})+ip_z(z-z^{\prime })\right] G(p)\Sigma _{-}.
\end{eqnarray}
Here $p=(p_0,0,0,p_z)$ and

\begin{equation}
G(p)=\left( \hat{p}+m_{e}\right) \left[ (p^{2}-m_{e}^{2}+i0)^{-1}+2\pi
i\delta (p^{2}-m_{e}^{2})N_{F}(p_{0})\right]
\end{equation}
is the Fourier transform of the Green's function in the two-dimensional
space (0, 3). For $N_{F}=0$ (no medium) the expression (15) is the
well-known, effectively two-dimensional, electron propagator used in the
theory of electrodynamic processes in superstrong magnetic fields and,
specifically, for investigation of the photon polarization operator.\cite
{r16}

{\bf 5}. Let us substitute the expression (15) into Eq. (3) and integrate
over $t,t^{\prime },y,y^{\prime },z,$ and $z^{\prime }$. This gives in the
form of a product of delta functions

\[
\prod_{n=0,y,z}\delta (k_n^{\prime }-k_n)\delta (p_n^{\prime }+k_n- p_n)
\]
the laws of conservation of energy and of the corresponding projections of
the momentum. The subsequent calculation of the Gaussian integrals over $x,$
$x^{\prime }$ and the trivial integral over $p_y$ gives $\delta (k_x^{\prime
}-k_x)$. As a result, as should be the case, we obtain a diagonal
representation of the polarization operator (5), where

\begin{equation}
\Pi (k)=\frac{G_a^2}\pi h\exp \left( -\frac{k_{\perp }^2}{2h}\right) \left[
F(l)+M(l)\right] ,
\end{equation}

\begin{equation}
F(l)=-i\int \frac{d^2p}{(2\pi )^2}T(l,p)\left[ p^2-m_e^2+i0\right]
^{-1}\left[ (p-l)^2-m_e^2+i0\right] ^{-1},
\end{equation}

\begin{equation}
M(l)=2\pi \int \frac{d^2p}{(2\pi )^2}\delta (p^2-m_e^2)N_F(p_0)\left[ \frac{%
T(l,p)}{(p-l)^2-m_e^2+i0}+(l\rightarrow -l)\right] .
\end{equation}
Here $p=(p_0,0,0,p_z)$ and $l=(k_0,0,0,k_z)\,$ are two-dimensional vectors,
and

\begin{equation}
T(l,p)=\frac 12\text{Tr\thinspace }\left[ \hat{k}\gamma ^5\left( \hat{p}%
+m_e\right) \Sigma _{-}\hat{k}\gamma ^5\left( \hat{p}- \hat{l}+m_e\right)
\Sigma _{-}\right] .
\end{equation}
In Eq. (17) the function $F$ corresponds to the purely field contribution,
and $M$ describes the influence of the medium. We note that $M$ does not
contain a term $\propto N_F(p_0)N_F(p_0-l_0)$, since

\[
\delta(p^2-m^2_e)\delta\left( (p-l)^2-m^2_e \right)\theta(p_0)
\theta(p_0-l_0)=0.
\]

Using the relations

\[
\left[ \Sigma _{-},\hat{p}\right] =0,\quad [\Sigma _{-},\gamma ^{5}]=0,\quad
\gamma ^{n}\Sigma _{-}=\Sigma _{+}\gamma ^{n}\quad (n=1,2),\quad \Sigma
_{+}\Sigma _{-}=0
\]
the trace of Eq. (20) reduces to a two-dimensional form and can be easily
calculated as

\begin{equation}
T(l,p)=\frac 14\text{Tr\thinspace }\left[ \hat{l}\left( \hat{p}+m_e\right)
\hat{l}\left( \hat{p}-\hat{l}-m_e\right) \right] =2(lp)^2- l^2(lp+p^2+m_e^2).
\end{equation}

We calculate the Gaussian integrals over $p_0$ and $p_z$ in Eq. (17) using
the trace (21) and the well-known Fock--Schwinger proper-time representation
for propagators of the form

\[
(\Delta +i0)^{-1}=-i\int\limits_{0}^{\infty }ds\,\exp \left[ is(\Delta
+i0)\right] .
\]
As a result, we find for the function $F(l)$ $\equiv F(l^{2},m^{2})$ the
integral representation

\begin{equation}
\overline{F}(l^{2})={-}i\frac{m_{e}^{2}\tau }{\pi }\int\limits_{0}^{1}dv\int%
\limits_{0}^{\infty }dx\left\{ \left[ 1{+}(1{-}v^{2})\tau \right] \exp
\left[ {-}ix\left[ 1{-}(1{-}v^{2})\tau \right] \right] {-}\exp ({-}%
ix)\right\} ,
\end{equation}

\[
\tau =l^{2}/4m_{e}^{2}.
\]
Here regularization is performed according to the well-known rule

\[
\overline{F}(l^{2})=F(l^{2},m^{2})-F(l^{2},\Lambda ^{2})
\]
with the regulator mass $\Lambda \rightarrow \infty $.

The renormalized polarization operator $\Pi _{R}(\tau )$ can be obtained
>from the dispersion relation with one subtraction (as well as it was done
for the photon polarization operator in Ref. 16)\vspace{0in}
\begin{equation}
\frac{1}{\tau }\Pi _{R}(\tau )=\frac{1}{\pi }\int\limits_{0}^{\infty }\frac{%
dt%
\mathop{\rm Im}%
\Pi \left( t\right) }{t\left( t-\tau -i0\right) }.
\end{equation}
Eq. (22) gives immediately
\begin{equation}
\mathop{\rm Im}%
\overline{F}=-\frac{l^{2}}{2}\int\limits_{0}^{1}dv\,\delta \left[
1-(1-v^{2})\tau \right] =-m_{e}^{2}\theta (\tau -1)\left( 1-\frac{1}{\tau }%
\right) ^{-1/2}.
\end{equation}
For $\tau <0$ we obtain from Eqs. (23), (24), and (17) the field contribution

\begin{equation}
\Pi _{R}^{(F)}=-\frac{\alpha _{a}}{\pi }m_{e}^{2}\frac{H}{H_{0}}\exp \left( -%
\frac{k_{\perp }^{2}}{2h}\right) \frac{1-\xi }{1+\xi }\ln \xi
\end{equation}
to the axion polarization operator. Here $\alpha _{a}=g_{ae}^{2}/4\pi $ (see
Eq. (4)), and the standard variable\cite{r17} $\xi $ was introduced as

\begin{equation}
\tau =-\frac{(1-\xi )^2}{4\xi },
\end{equation}
which is convenient for analytical continuation in $l^2=4m_e^2\tau $.

For $\tau >1$, a channel is open for axion decay into an electron-positron
pair $(a\rightarrow e^{-}e^{+})\,$ in a magnetic field. Its rate $w$ for a
real axion is related with the imaginary part of the polarization operator
on the mass shell by the well-known relation

\begin{equation}
w=-\frac 1\omega
\mathop{\rm Im}%
\Pi _R^{(F)}=\alpha _a\frac{m_e^2}\omega \,\frac H{H_0}\exp \left( - \frac{%
k_{\perp }^2}{2h}\right) \theta (\tau -1)\left( 1-\frac 1\tau \right)
^{-1/2},
\end{equation}
where $\omega $ is the axion energy.

This result, which follows from Eq. (25) with $\xi =\left| \xi \right| \exp
(i\pi )\,$ (see Eq. (26)), is identical to the result obtained in Ref. %
\onlinecite{r13} on the basis of a calculation of the elastic scattering
amplitude of an axion in a magnetic field. It can also be found at once from
Eqs. (24) and (17).

Let us consider the contribution $M$ (19) of the medium to the axion
polarization operator. We note that it does not renormalize.\cite{r15}
Integrating over $p_{z}$ in Eq. (19), using the delta function and taking
account of Eqs. (13) and (21), gives

\[
M=-\frac{m_e^2}{2\pi }l^2\int\limits_{m_e}^\mu \frac{d\varepsilon }q\left[
D(l,p)+D(-l,p)+D(l,\tilde{p})+D(-l,\tilde{p})\right] ,
\]
\begin{equation}
D(l,p)=\left[ l^2-2(lp)+i0\right] ^{-1}.
\end{equation}
Here $\varepsilon $ is the energy of electrons in the medium, $q=\sqrt{%
\varepsilon ^2-m_e^2}$, and the two-dimensional scalar products are $%
lp=k_0\varepsilon -k_zq$ and $l\tilde{p}=k_0\varepsilon +k_zq$.

The imaginary part of the expression (28) is determined using Sokhotski\u{\i}%
's formula

\begin{equation}
\frac{1}{x+i0}={\rm P}\frac{1}{x}-i\pi \delta (x),
\end{equation}
where P signifies a principal value. From Eqs. (28) and (29) we obtain on
the mass shell

\[
\mathop{\rm Im}%
M=\frac{m_e^2}2\theta (\tau -1)\left[ \theta (\mu -\varepsilon _{+})+\theta
(\mu -\varepsilon _{-})\right] ,
\]
\begin{equation}
\varepsilon _{\pm }=\frac \omega 2\pm \frac{k_z}2\left( 1-\frac 1\tau
\right) ^{1/2}.
\end{equation}
Here $\varepsilon _{\pm }$ are the roots of the equations $l^2- 2\omega
\varepsilon \pm 2k_zq=0$.

>From Eqs. (17), (24), and (30) we find the rate

\begin{equation}
w_{M}=\frac{1}{2}\left[ \theta (\varepsilon _{+}-\mu )+\theta (\varepsilon
_{-}-\mu )\right] w,
\end{equation}
where $w$ is the decay rate (27) in the absence of a medium, for the axion
decay into an $e^{-}e^{+}$ pair in the presence of a magnetized degenerate
electron gas. We underscore that the imaginary part of the contribution (30)
of the medium is positive, and summed with the negative field contribution
(24) it gives a blocking Pauli factor $1-\theta (x)=\theta (-x)$ in Eq.
(30). It forbids electron production inside a filled Fermi sphere (for $%
\varepsilon _{\pm }<\mu $).

Taking account of Eq. (29), we obtain for the real part of Eq. (28) on the
mass shell the representation
\begin{equation}
\mathop{\rm Re}%
M=-\frac{m_{e}^{2}}{\pi }\tau
\mathop{\diagup\!\llap{$\displaystyle \!
               \int\limits_{0}^{\lambda} $}}dx\left[ \frac{1}{\tau -\cosh
^{2}(x-\psi )}+\frac{1}{\tau -\cosh ^{2}(x+\psi )}\right] .
\end{equation}
Here the substitution of the variable $\varepsilon \rightarrow x$ was used: $%
\varepsilon =m_{e}\cosh x$ and $q=m_{e}\sinh x$, and the parameters $\lambda
$ and $\psi $, defined as

\begin{equation}
\cosh \lambda =\frac{\mu }{m_{e}},\quad \tanh \psi =\frac{k_{z}}{\omega },
\end{equation}
were introduced. The integral (32) can be expressed in terms of elementary
functions.

We shall confine our attention below to the limiting cases that are of
interest for astrophysical applications.

{\bf 6}. For an axion on the mass shell

\begin{equation}
l^{2}=4m_{e}^{2}\tau =\omega ^{2}-k_{z}^{2}=m_{a}^{2}+k_{\perp }^{2}>0,
\end{equation}
and the condition (14) gives $k_{\perp }^{2}\ll h$, so that $\exp (-k_{\perp
}^{2}/2h)\simeq 1$. We note that the imaginary part of the polarization
operator is formed by the contribution of real electrons and positrons, and
the expression for it holds under the weaker condition $k_{\perp }^{2}<2h$.
Therefore the exponential factor can be retained in Eq. (27).

For $\tau \ll 1$ (substantially below the threshold of the decay process $%
a\rightarrow e^{-}e^{+}$), we find from Eqs. (25), (32), and (17)

\begin{equation}
\Pi _{R}=\Pi _{R}^{(F)}+\Pi ^{(M)}=-\frac{\alpha _{a}}{\pi }m_{e}^{2}\frac{H%
}{H_{0}}\tau \left( 2-\nu _{+}-\nu _{-}\right) .
\end{equation}
Here

\[
\nu _{\pm }=\tanh (\lambda \pm \psi )=\frac{\nu \omega \pm k_{z}}{\omega \pm
\nu k_{z}},\quad \nu =\tanh \lambda =\left[ 1-\left( \frac{m_{e}}{\mu }%
\right) ^{2}\right] ^{1/2}.
\]
We note that $\Pi _{R}<0$ and if the axion moves in the direction of the
field ${\bf H}$ $(k_{\perp }=0)$, then according to Eqs. (34) and (35) $\Pi
_{R}\rightarrow 0$ in the limit of a massless axion $(m_{a}\rightarrow 0)$.

At high energies $(\tau \gg \left( \mu /m_{e}\right) ^{2}\gg 1)$ we obtain
for the polarization operator the asymptotic representation

\begin{equation}
\Pi _{R}=\frac{\alpha _{a}}{\pi }m_{e}^{2}\frac{H}{H_{0}}\left[ \ln \left[
\tau \left( \frac{m_{e}}{\mu }\right) ^{2}\right] -i\pi \right] \simeq \frac{%
2\alpha _{a}}{\pi }eH\left[ \ln \frac{k_{\perp }}{2\mu }-i\frac{\pi }{2}%
\right] ,
\end{equation}
and it does not depend of the electron mass $m_{e}$ as it should be in this
limit.

Let us write the dispersion relation (7) in the form

\begin{equation}
\omega ^{2}=k_{\perp }^{2}+k_{z}^{2}+m_{a}^{2}+\Pi _{R}(k).
\end{equation}
It follows from Eqs. (35) -- (37) that in a magnetized medium a radiative
shift of the axion mass is generated --- a dynamic mass, whose square,
according to the definition in Ref. \onlinecite{r15}, is

\[
\delta m_a^2=%
\mathop{\rm Re}%
\Pi _R.
\]

For $\tau \gtrsim 1$ and $\mu /m_{e}\gg 1$ we obtain the estimate

\begin{equation}
\delta m_{a}\sim g_{ae}m_{e}\left( \frac{H}{H_{0}}\right) ^{1/2}\sim
10^{6}g_{ae}\left( \frac{H}{10^{13}\text{ G}}\right) ^{1/2}\text{{\rm \ }eV}.
\end{equation}
For $g_{ae}\sim 10^{-13}$ (Refs. \onlinecite{r4} and \onlinecite{r10}) and $%
H\gtrsim 10^{17}$ G (such fields\cite{r18,r19} and even $H\sim
10^{18}-10^{20}$ G (Ref. \onlinecite{r20}) can exist in the interior regions
of neutron stars), Eq. (38) gives $\delta m_{a}\gtrsim 10^{-5}$ eV.

The chemical potential $\mu $ of a degenerate gas of electrons occupying the
ground-state Landau level $(n=0)$ in a magnetic field is related with the
electron density $n_e$ by the well-known relation

\begin{equation}
n_e=\frac{hp_F}{2\pi ^2},
\end{equation}
where $p_F=\sqrt{\mu ^2-m_e^2}$ is the Fermi momentum. Writing Eq. (12) in
the form

\begin{equation}
\frac H{H_0}>\frac 12\left( \frac{p_F}{m_e}\right) ^2,
\end{equation}
we obtain, taking account of Eq. (39), an upper limit on the density

\begin{equation}
n_e<\frac{\lambdabar_e^{-3}}{\sqrt{2}\,\pi ^2}\,\left( \frac H{H_0}\right)
^{3/2},
\end{equation}
where $\lambdabar_e=1/m_e$ is the electron Compton wavelength. For $H=2\cdot
10^{17}$ G Eqs. (40) and (41) give $p_F<50\,$ MeV and $n_e<10^{36}$ cm$^{-3}$%
. Next, let $T\sim 10^{10}$ K $\sim 1$ MeV and $k_{\perp }\gtrsim T$. Then
the conditions (11), (12), and (14) can be satisfied and the estimate (38)
can be justified.

In summary, under the conditions of strongly magnetized neutron stars a
dynamic axion mass, which can fall within the existing limits on the axion
mass\cite{r3,r4,r12} --- $10^{-5}$ eV $\lesssim m_{a}\lesssim 10^{-2}$ eV
--- is generated. Therefore $\delta m_{a}\sim m_{a}$ and the dispersion
relation (37) differs appreciably from the vacuum relation $%
(k^{2}=m_{a}^{2}) $. This must be taken into account, for example, when
investigating the resonant conversion of a plasmon into an axion $(\gamma
\rightarrow a)$ in a magnetic field as a result of the crossing of the
corresponding dispersion curves (as already noted above, this process in
fields $H\ll H_{0}$ and in the absence of the direct coupling (1) was
studied in Ref. \onlinecite{r12}). We also note that the rate (27) of the
decay $a\rightarrow e^{-}e^{+}\,$ in a magnetic field has a square-root
threshold singularity (as $\tau \rightarrow 1+0$). This singularity can be
removed by taking into account accurately the dispersion law of an axion
near threshold, and the decay rate is found to be finite:\cite{r13} $w\sim
m_{e}(\alpha _{a}H/H_{0})^{2/3}$. A detailed analysis of the same threshold
singularity (of cyclotron resonance) in a magnetic field and its elimination
for the photon decay process $(\gamma \rightarrow e^{-}e^{+})$ was given
earlier in Ref. \onlinecite{r21}, where, specifically, it is underscored
that the indicated singularity can be explained by the quantization of the
phase space of charged particles in a magnetic field.

Translated by M. E. Alferieff

\end{document}